\documentclass[twocolumn,prb,aps,superscriptaddress,showpacs]{revtex4}
\usepackage{graphicx}
\begin{document}
\title {Target-searching on the percolation}
\author{Shi-Jie Yang}
\affiliation{Department of Physics, Beijing Normal University,
Beijing 100875, China}
\date{\today}

\begin{abstract}
We study target-searching processes on a percolation, on which a
hunter tracks a target by smelling odors it emits. The odor
intensity is supposed to be inversely proportional to the distance
it propagates. The Monte Carlo simulation is performed on a
2-dimensional bond-percolation above the threshold. Having no idea
of the location of the target, the hunter determines its moves
only by random attempts in each direction. For lager percolation
connectivity $p\gtrsim 0.90$, it reveals a scaling law for the
searching time versus the distance to the position of the target.
The scaling exponent is dependent on the sensitivity of the
hunter. For smaller $p$, the scaling law is broken and the
probability of finding out the target significantly reduces. The
hunter seems trapped in the cluster of the percolation and can
hardly reach the goal.
\end{abstract}
 \maketitle

\section{Introduction}
In the past years, diffusion-controlled reactions have been
extensively studied through random-walk models. Such applications
range from chemical processes, electronic scavenging and
recombination, to electronic and vibrational energy transfer in
condensed
media\cite{Montroll,Klafter,Stanley,Szabo,Koza,Bere,Agmon}. Many
works have been devoted to the target annihilation problem, in
which randomly placed targets are annihilated by random walkers,
and its dual of the trapping problem\cite{Weiss2,Jasch}. Other
models treat hindered diffusion problems which involve random
point obstacles \cite{Saxton1}. In these models, the tracer moves
from site to site on a lattice and falls into wells of various
depth at the sites. Another possibility is a mountain model, in
which all sites are at zero energy and the barriers are on the
bonds joining the sites\cite{Saxton3}. In the bond percolation
problem, all sites are, in principle, accessible but the lattice
is restricted by labelling the bonds which connect two sites as
closed or open with probability $p$ and $1-p$,
respectively\cite{Stauffer}. There exists a critical probability
$p_c$ (percolation threshold) such that an infinite conducting
cluster exists for $p>p_c$ and does not exist for $p<p_c$. The
percolation threshold depends on the geometric properties of the
lattice and the kind of percolation problem to be studied (site or
bond). For the bond percolation, $p_c=0.5$ exactly in two
dimensions and $p_c\approx 0.2488$ in three
dimensions\cite{Isichenko}.

In this paper, we present the target-searching game on a
two-dimensional bond percolation. The active hunter is trying to
find out a target which emits a special kind of odor. This kind of
hunting processes, which frequently occurs in biological systems,
such as a shark searching for foods by smelling the blood in the
ocean, or honeybees flying in the countryside to locate the
foraging-nectars\cite{Collett,Kareiva,Marsh}, or in metabolic
processes such as cell motions and
chemotaxis\cite{Alt,Keller,Mah,Shenderov,Sherratt}, can be viewed
as the target-oriented problems. The hunters try to reach the
targets by following some kind of behavior rules. The structure of
percolation can properly reflect the randomness of the
environment.

The Monte Carlo simulations are performed on a 2-dimensional bond
percolation on an underlying square lattice. Since neither the
distance nor the direction of the target is presumedly to be
known, the searcher should determine its moves by random attempts
in each direction, just like a snake turns its head from side to
side to test the variation of the odor intensity. There is some
chance for the hunter to move in the wrong direction because of
randomness. Hence it is not a traditional biased random-walk.
After sorting each searching process in a time sequence by
introducing a variable $x$ to represent as a percentage, we find a
scaling law for the searching time versus the distance to the
position of the target for percolating probability $p\gtrsim
0.90$. The scaling exponent is found to be dependent on the
sensitivity of the hunter. For $p\lesssim 0.90$, the scaling law
is broken and the probability of reaching the goal reduces. The
vacancy of some bonds frustrates the normal searching process.
Under some circumstance, the hunter may be trapped in dead-end
branches of the percolating cluster which it can hardly get out
because of the constraint from the game rules.

The paper is arranged as following: The model is described in
section II. The scaling property for larger bond connectivity is
studied in section III and the trapping effect for smaller
connectivity is studied in section IV. In section V, we discuss
the trapping property and compare it with the trapping effect in
biased random walks. A summary is included in section VI.

\section{The model}
In our Monte carlo experiment, we prepare a $3000\times 3000$
underlying regular lattice and randomly remove a fraction $1-p$ of
the bonds. We restrict us to $p>p_c$ so that the network forms an
infinite percolating cluster. The game rules are as follows:  The
hunter at the origin $O$ is trying to find out a target somewhere
which emits a special kind of odor. Since the hunter have no way
to know the location of the target, it randomly moves around its
present position to test the variation of the odor intensity.
$z_0$ is the present distance of the hunter to the target while
$z_1$ is the corresponding distance of the next attempted step.
Here distance is measured as the Euclidean distance. The Monte
Carlo steps are implemented as: if $(z_0/z_1)^\alpha>\zeta$, where
the parameter $\alpha$ reflects the sensitivity of the hunter and
$\zeta$ is a random number, and there is a bond linking the two
sites, then the attempt is accepted. Otherwise it is refused. This
rule implies that the intensity of the signal emitted by the
target is inversely proportional to the distance of the hunter to
the target. Other choices of the relation do not alter the result
qualitatively. By this way, the hunter approaches the goal in a
stochastic style. When the hunter gets the goal, this round of
search is over. The underlying lattice is large enough so that the
hunter will not go beyond the edges of the percolation in the
course of searching.

A total of $2000$ simulations are performed for each given
distance to the target. When a round of simulation is over, we
record the searching time $t$. Then the time of all simulations
are resorted increasingly by introducing a variable $x$ to
represent the percentage of searching times under a given value.
When the hunter is far away from the goal, the ratio $z_0/z_1$ is
close to $1$. Most of the moving attempts are accepted, even the
hunter walks in the wrong direction. The hunter appears to linger
around for quite a while. Hence the motion of the hunter is nearly
a Brownian random walk. As the goal is nearer, the ratio of
$z_0/z_1$ gradually approaches $0.5$ and the probability of being
refused for the hunter moving in the wrong direction increases.

Figure 1(a) is a log-normal distribution of searching time $t$
versus $x$ for an initial distance $z=137.1$ to the target. The
sensitivity parameter $\alpha=6$. Each curve represents a
different bond-connected fraction $p$ ranging from $0.60$ to
$0.97$. In this figure, the curves are divided into two distinct
regimes. In the smaller connectivity regime, the searching time
$t$ increases with reduction of $p$, indicating that the
percolation frustrates the searching process. In the larger
connectivity regime ($p\gtrsim 0.90$), the searching times almost
keep invariant, indicating that small randomness can hardly affect
the move of the hunter. Figure 1(b) displays the relation of the
searching time and percolation connectivity $p$ for $x=50$. The
smaller for $p$, the longer for searching time. Below we will
study the properties of the two regimes.

\section{The scaling regime}
Figure 2(a) is a log-normal plot of the searching time $t$ versus
$x$ for different initial distances at a percolation probability
$p=0.95$, with $z=31.4,65.6,137.1,188.4,350.0$ from lower to upper
and $\alpha=6$. These curves are well-parallel to each other.
Figure 2(b) shows linear relations of $\ln t$ and $\ln z$ at
$x=50$. From upper to lower, $\alpha=6,10,14,18,22$, respectively.
Hence each curve can be described by a single function
$f(x,\alpha)$, which reflects the statistical distribution of the
searching time in the each simulation, plus a $z$-dependent
function,
\begin{equation}
\ln t(x,z,\alpha)=f(x,\alpha)+\eta (\alpha)\ln z. \label{fz}
\end{equation}
In Fig. 3 we study the dependence of the searching-time with
respect to the sensitivity parameter $\alpha$. Figure 3(a) shows
the curves for $\alpha=6,10,14,18,22$, respectively. After
properly rescaling the curves in (a) by times $\ln t$ with a
coefficient $\alpha^\beta$, all the curves become parallel
(Fig.3(b)). From Fig.3(b), we deduce $\alpha^\beta \ln t=\tilde f
(x)+\tilde \phi (z,\alpha)$, where $\beta=0.623$. By comparing
with Eq.(\ref{fz}), one gets
\begin{equation}
\ln t(x,z,\alpha)=\alpha^{-\beta} f(x)+\eta (\alpha)\ln z.
\end{equation}
To go further, we probe the relation of searching with parameter
$\alpha$, as shown in Fig.4. It reveals a good linear relation
$\ln (\alpha^{\beta}\ln t)\propto \gamma \ln \alpha$, with the
slope $\gamma=0.5$.

Combining all the above factors, we can figure out a scaling law
for the searching-time with respect to the distance as well as
sensitivity parameter $\alpha$,
\begin{equation}
\ln t=\alpha ^{-\beta}f(x)+\alpha^{\gamma-\beta}\ln z,
\end{equation}
or
\begin{equation}
t(x,z,\alpha)\propto e^{f(x)/\alpha^{\beta}}\cdot
z^{\delta}.\label{fr}
\end{equation}
The scaling exponent is found to be $\delta=
\alpha^{\gamma-\beta}$. It shows that the more sensitive the
hunter, the less time it costs to reach the target.

\section{The trapping regime}
Now we turn to study the properties of the regime $p<0.90$. As
shown in Fig.1, the searching processes for $p<0.90$ are greatly
affected by the absence of the bonds, with larger searching time
for smaller fraction of bond-connectivity $p$. This means that
strong disorders tend to block the searching process of the
hunter. Furthermore, the scaling law revealed in section III is
broken.

Figure 5 shows the crossover from the scaling regime to the
trapping regime. The search processes are averaged over 200
configurations of percolation. Given the upper limit of walking
steps $t_c$, above which one can consider that the hunter fails to
reach the target. The vertical axis in Fig.5 is the probability of
success to find out the target within the time period $t_c$. In
Fig.5(a), the initial distance $z=137.1$ and $\alpha=6,10,14,18$
from upper to lower. $t_c=1,000,000$ steps for all simulations. It
can be seen that for $p>0.90$, the hunter always reaches the goal
in this time limit. As $p$ decreases, the probability of missing
the goal increases significantly. It is notable that the more
sensitive, the more probable for the hunter to fail to reach the
goal. This trend is just reverse with that in the scaling regime.
Under some circumstance, the hunter will never reach the goal
because the searching time goes to infinity. The hunter seems
trapped in some dead-end branches of the percolation, although the
percolation is still well-connected. We call this range of
connectivity $p$ the trapping regime. Figure 5(b) is the
probability of success for three different time limits
$t_c=100,000, 500,000, 1,000,000$, with $\alpha=6$. Not
surprisedly, the probability of success increases with the upper
limit of walking time $t_c$.

\section{Discussion}
It should be mentioned that the trapping of the hunter in the
homogenous percolating cluster is somehow analogous to the
trapping of a random walker in a strong biased diffusion. When the
biased field $B$ exceeds a critical value $B_c$, the walkers may
enter a dead-end branch so that it is difficult to escape away by
overcoming the biased field\cite{Dhar,Stauffer2,Reyes}. The drift
velocity tends to zero and the particle will never reach the
opposite end. This is easily seen: For a trap of depth $l$ the
potential barrier to cross increases with $l$, and the trapping
time varies as $(\frac{1+B}{1-B})^l$. The density $\rho(l)$ of
traps of depth $l$ varies as $\exp(-l/\xi)$, where $\xi$ is the
$p$-dependent percolation correlation length of the system. Hence
the average trapping time per step along the backbone is
\begin{equation}
\sum_{l=1}^{\infty}\rho (l)(\frac{1+B}{1-B})^l
\end{equation}
This summation converges only for $B<B_c=tanh(\frac{1}{2\xi})$.
For $B>B_c$, the asymptotic velocity is zero, but the mean
displacement of the particle $\langle z \rangle$ increases as
$t^a$ with $a<1$. In a time $t$, the particle can, on the average,
only travel a distance $t^a$ before it encounters a trap with
trapping time bigger than $t$, and gets stuck there. Eventually,
it will exit from this trap, only to get stuck in other traps. In
our case, however, the constrain come from the hunter trying to
trace the target. When it unfortunately enters a series of
dead-end branches of the infinite cluster, it will be difficult to
get out of them. In the biased diffusion processes, the larger
biased field leads to more strict trapping. In the hunting
processes, increase $\alpha$ will also enforce the trapping
effect.

\section{Summary}
We have studied the searching processes on a percolation. It is
found that when de disorders are low, there is a scaling law
between the searching time and the distance to the target. For
strong disordered systems, the scaling law is broken. The hunter
may be trapped in the dead-end branches and can hardly reach the
target. This trapping effect is somehow analogous to the trapping
in the biased random walk processes. Our study may be instructive
for zoologists or entomologists when they explore the
food-foraging behavior of wild animals.

\centerline {Figure Captions}

Figure 1 (a) Time-sorted curves for 2000 simulations. The
horizontal axis $x$ is the sequence of searching time represented
as a percentage and the vertical axis is logarithmic time.
$\alpha=6$. From upper to lower,
$p=0.6,0.7,0.8,0.9,0.93,0.95,0.97$. As $p\gtrsim 0.90$, the curves
almost fall into one. (b) $p$-dependence of searching time for
$x=50$.

Figure 2 (a) Time-sorted curves at $p=0.95$ for various distances
$z$ from the origin, with $z=31.4,65.6,137.1,188.4$ from bottom to
top. These curves are parallel to each other. (b) A plot of $\ln
t$ versus $\ln z$ at $x=50$ for various sensitivity parameter.
From upper to lower, $\alpha=6,10,14,18,22$, respectively. It
reveals a linear relation.

Figure 3 Time-sorted curves for 2000 simulations for various
parameter values $\alpha$. The horizontal axis $x$ is the sequence
of searching times represented as a percentage and the vertical
axis is logarithmic time. The original distance is fixed at
$z=137.1$. (a) is for $\alpha=6,10,14,18,22$ from top to bottom.
(b) shows the rescaled curves of (a) for $\alpha=6,10,14,18,22$
from bottom to top. These curves are parallel to each other.

Figure 4 A linear relation of $\ln (\alpha^\beta \ln t)$ vs $\ln
\alpha$ at $x=50$. The distance $z=137.1$. The slope is
$\gamma=0.5$.

Figure 5 (a) Crossover from the scaling regime to trapping regime.
The vertical axis is the probability of success of finding out the
goal at a given time period $t_c=1,000,000$. The horizonal axis is
the bond-connected fraction. The original distance $z=137.1$. (b)
The same as in (a) for various time limits. The sensitivity
parameter is fixed at $\alpha=6$.

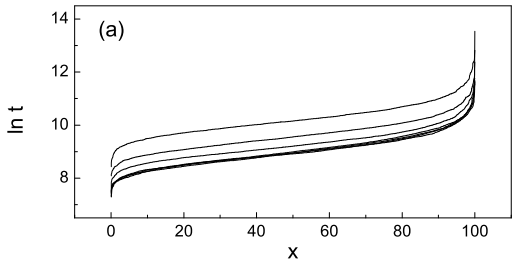
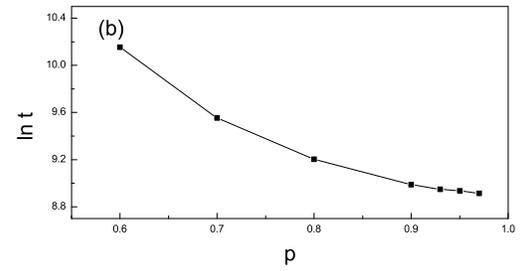
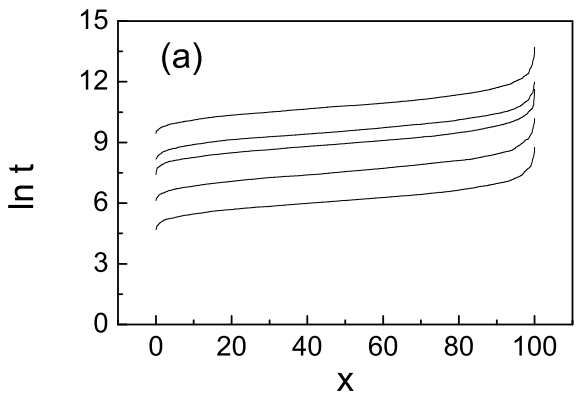
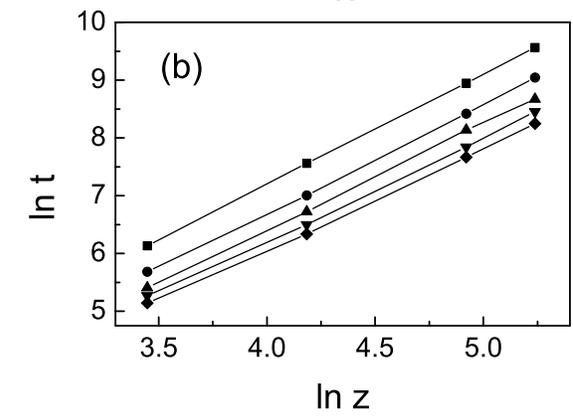
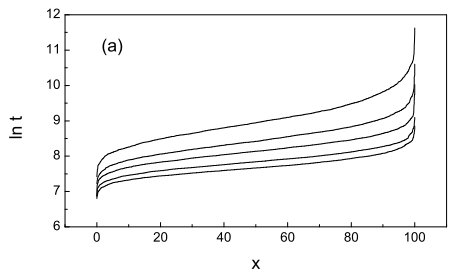
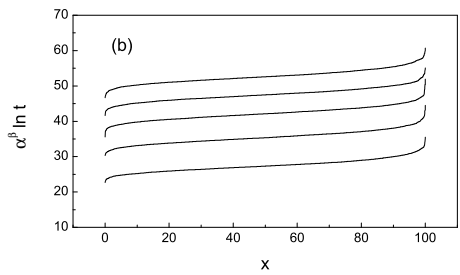
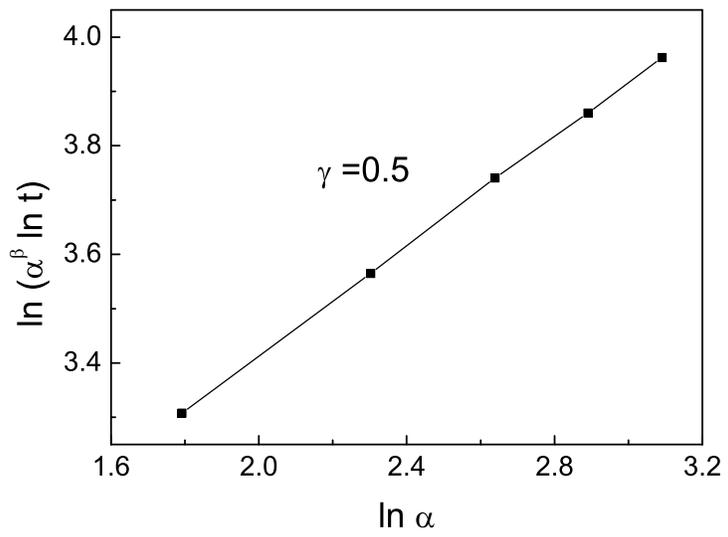
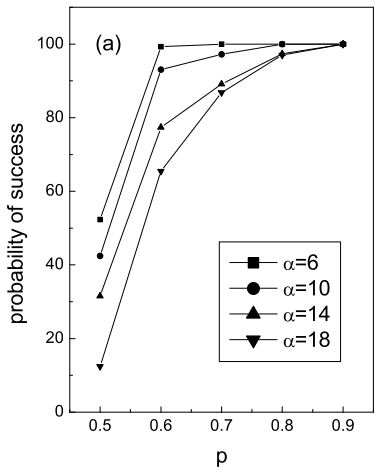
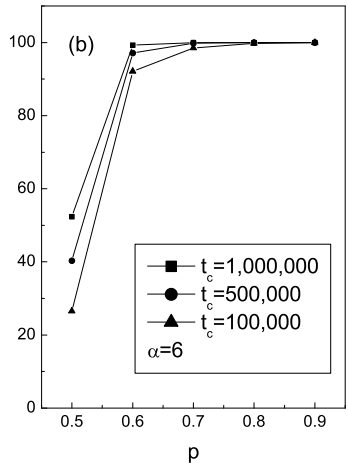
\end{document}